\documentclass[reprint,aps,prb,showpacs]{revtex4-1}
\usepackage{color,graphicx,calc,url}
\usepackage{dcolumn}
\usepackage{bm}
\usepackage{amssymb}
\usepackage{amsmath}
\usepackage{wasysym}
\usepackage{natbib}
\usepackage{mathrsfs}

\begin{document}
\title{Comparison of laser induced and intrinsic tunnel magneto-Seebeck effect in CoFeB/MgAl$_2$O$_4$ and CoFeB/MgO magnetic tunnel junctions}
\author{Torsten Huebner$^1$, Alexander Boehnke$^1$, Ulrike Martens$^2$, Andy Thomas$^3$, Jan-Michael Schmalhorst$^1$, G\"unter Reiss$^1$, Markus M\"unzenberg$^2$,  Timo Kuschel$^{1,4}$ \email{Electronic mail: thuebner@physik.uni-bielefeld.de}}
\affiliation{$^1$Center for Spinelectronic Materials and Devices, Department of Physics, Bielefeld University, Universit\"atsstra\ss e 25, 33615 Bielefeld, Germany\\
$^2$Institut f\"ur Physik, Greifswald University, Felix-Hausdorff-Strasse 6, 17489 Greifswald, Germany\\
$^3$Institute for Metallic Materials, IFW Dresden, Helmholtzstra\ss e 20, 01069 Dresden, Germany\\
$^4$Physics of Nanodevices, Zernike Institute for Advanced Materials, University of Groningen, Nijenborgh 4, 9747 AG Groningen, The Netherlands}

\date{\today}


\begin{abstract}
We present a comparison of the tunnel magneto-Seebeck effect for laser induced and intrinsic heating. Therefore, Co$_{40}$Fe$_{40}$B$_{20}$/MgAl$_2$O$_4$ and Co$_{25}$Fe$_{55}$B$_{20}$/MgO magnetic tunnel junctions have been prepared. The TMS ratio of 3\,\% in case of the MAO MTJ agrees well with ratios found for other barrier materials, while the TMS ratio of 23\,\% of the MgO MTJ emphasizes the influence of the CoFe composition. We find results using the intrinsic method that differ in sign and magnitude in comparison to the results of the laser heating. The intrinsic contributions can alternatively be explained by the Brinkman model and the given junction properties. Especially, we are able to demonstrate that the symmetric contribution is solely influenced by the barrier asymmetry. Thus, we conclude that the symmetry analysis used for the intrinsic method is not suitable to unambiguously identify an intrinsic tunnel magneto-Seebeck effect.
\end{abstract}

\maketitle
\section{Introduction}
Spin caloritronics is a rising field of research seeking to combine spin, charge, and temperature driven currents to develop new and improved ways of data processing and storage. Especially, the usage of temperature driven spin-polarized currents has attracted a lot of attention in the past years, since it may offer a way to deal with rising heat dissipation in nano devices \cite{Bauer}.

Lately, significant progress has been made in controlling temperature differences in magnetic nanostructures over a small distance \cite{nature}, enabling the discovery of, for example, the \textit{tunnel magneto-Seebeck} (TMS) effect \cite{walter,Liebing1}. This effect occurs in \textit{magnetic tunnel junctions} (MTJs) when a temperature difference is generated across the barrier. The TMS effect describes the change of the Seebeck coefficient ($S_\text{p}$ and $S_\text{ap}$) of the MTJ between the state of parallel ($p$) and antiparallel ($ap$) relative magnetization orientation of the ferromagnetic electrodes. The effect ratio \cite{czerner} can be expressed by
\begin{align}\label{eq:TMS}
\text{TMS} = \frac{S_{\text{p}}-S_{\text{ap}}}{\text{min}(\left|S_{\text{p}}\right|,\left|S_{\text{ap}}\right|)}\ \ \ .
\end{align}

Recently, the reciprocal effect, the magneto-Peltier effect, has also been reported for MTJs \cite{magnetopeltier}.
Today, different heating methods are established to generate a temperature difference inside the MTJ: indirect Joule \cite{Liebing1,Liebing2,Liebing3}, indirect Peltier \cite{magnetopeltier}, and laser induced heating \cite{walter,Boehnke1,Boehnke2}. An additional method proposes to use the direct intrinsic Joule heating by the tunneling current. With this method, the temperature difference is created without additional external heating and, thus, the effect is called the \textit{intrinsic} TMS effect \cite{Zhang,Teixeira}.

In most cases CoFeB/MgO/CoFeB MTJs are used to study the TMS effect, because they are well known, easily prepared and show large \textit{tunnel magnetoresistance} (TMR) effects \cite{drewello}. Using MgAl$_2$O$_4$ (MAO) as a barrier material theoretically retains the aforementioned properties of MgO (for example the $\Delta_1$ symmetry filter effect \cite{meff}) whereas the lattice mismatch with typical electrode materials decreases from about (3-5)\,\% for MgO to about 1\,\% for MAO \cite{miura}. In addition, MgAl$_2$O$_x$ double-barrier MTJs show a long-range phase coherence using the resonant states of Fe quantum wells with up to 12\,nm thickness \cite{tao}. Here, the structural flexibility of MgAl$_2$O$_x$ ensures a vanishing mismatch between barrier and electrode, effectively enhancing quantum phenomena.
Additionally, an improved bias voltage dependence was found with a barrier consisting of MAO \cite{Sukegawa1}. A maximum TMR ratio of over 160\,\%, a very low \textit{resistance area product} ($RA$) of less than 5\,$\Omega\mu$m$^2$, as well as magnetization switching by spin-transfer torque was achieved by depositing and oxidizing Mg/Mg-Al layers \cite{Sukegawa2}. 

In this work, we study CoFeB/MAO and CoFeB/MgO MTJs and place emphasis on the comparison of laser induced and intrinsic TMS. After a description of the sample preparation in Sec. \ref{prep}, the results of the TMR and laser induced TMS measurements are presented in Sec. \ref{results}, followed by COMSOL simulations of the temperature differences, the Brinkman model and the results of the intrinsic TMS.  

\section{Sample preparation}\label{prep}

The CoFeB/MAO and CoFeB/MgO layer stacks are deposited on MgO(001) substrates to prevent parasitic effects originating from semiconducting substrates as reported in Ref. \onlinecite{Boehnke1}. The sequence of layers of the MAO MTJ consists of a bottom contact Ta 10/Ru 30/Ta 5/Ru 5, a pinned layer MnIr 10/Co$_{40}$Fe$_{40}$B$_{20}$ 2.5, a tunnel barrier MAO 1.8, a free layer Co$_{40}$Fe$_{40}$B$_{20}$ 2.5 and a top contact Ta 5/Ru 30/Ta 5/Au 60 (numbers are thicknesses in nm).  Except for MAO, all films are prepared by \textit{dc} sputtering at a base pressure of less than $5\cdot10^{-7}$\,mbar in a Leybold Vakuum GmbH CLAB 600. MAO is \textit{rf} sputtered from a composite target in the same chamber. After deposition, ex-situ post annealing in a vacuum furnace is carried out at 350\,$^{\circ}$C for one hour with a subsequent cooling process in a magnetic field of 0.7\,T. Elliptical junctions of 24\,$\mu$m$^2$ are prepared by electron beam lithography and subsequent ion beam etching. Ta$_2$O$_5$ (120\,nm) is sputtered next to the MTJs to serve as insulator. In addition, Au bond pads are placed on top and next to the MTJs to allow electrical contacting via Au bonds and optical access. 
The experimental details of the Co$_{25}$Fe$_{55}$B$_{20}$ 2.5/MgO 1.7/Co$_{25}$Fe$_{55}$B$_{20}$ 5.4 (numbers are thicknesses in nm) MTJ are described elsewhere \cite{Boehnke2}.
To measure the TMS effect, an established setup with a modulated diode laser (P$_{\text{max}}$=150\,mW, $\lambda$=637\,nm, $f$=177\,Hz) is used to generate the temperature difference across the junction (see Refs. \onlinecite{ walter} or \onlinecite{ Boehnke1} for details). At the same time, the setup is able to record TMR loops and I/V characteristics with a \textit{Keithley 2400 Sourcemeter}.

\section{Results}\label{results}

\subsection{TMR and laser induced TMS}
\begin{figure}[bt]\centering
		\includegraphics{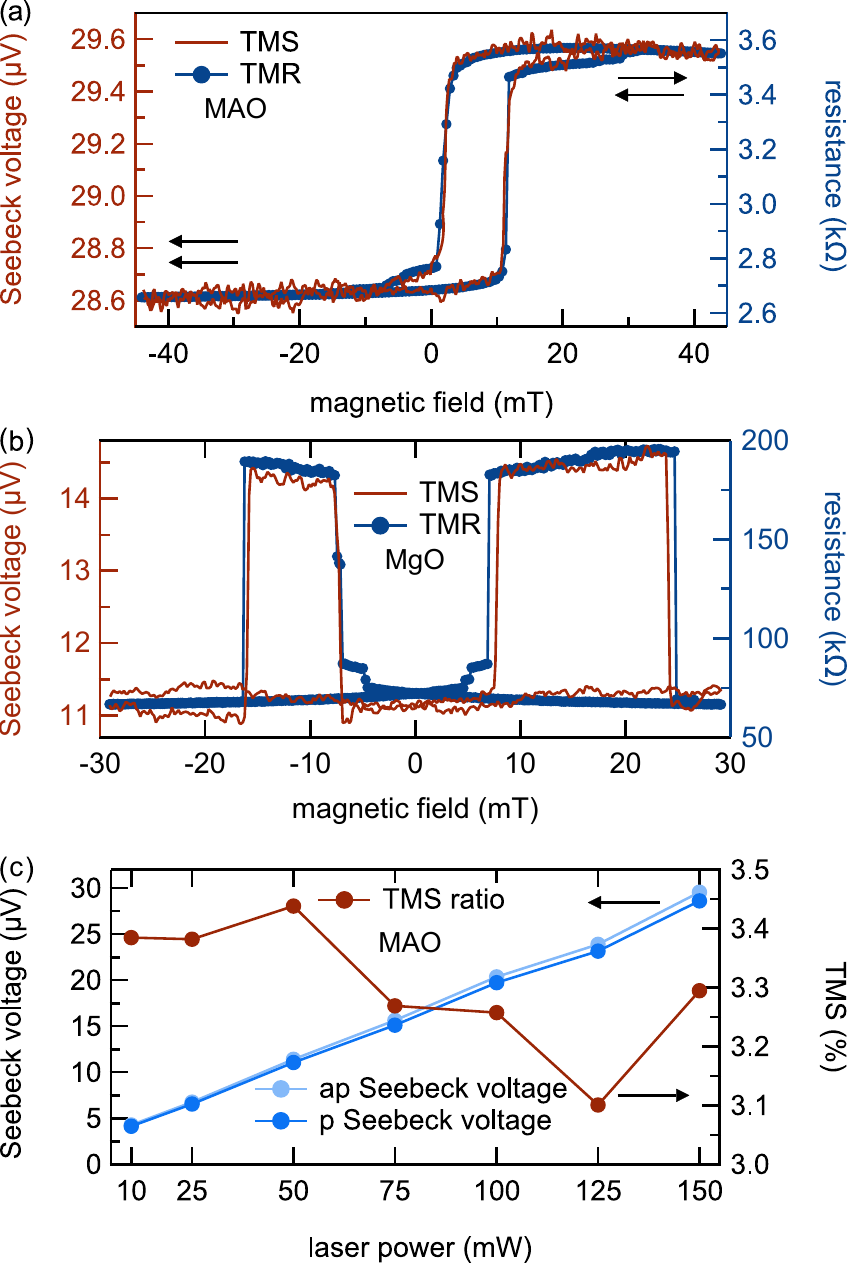}
	\caption{(a) Typical TMS (measured with a laser power of 150\,mW) and TMR minor loop of a junction with an area of 24\,$\mu$m$^2$  and a nominal MAO thickness of 1.8\,nm. (b) TMS (150\,mW) and TMR major loops of an CoFeB/MgO MTJ with a junction size of 2\,$\mu$m$^2$ and a nominal MgO thickness of 1.7\,nm. (c) Antiparallel (light blue) and parallel (dark blue) Seebeck voltages for an MTJ with an MAO thickness of 1.8\,nm increase linearly with the laser power, while the TMS ratio (red) is constant ($3.3\pm0.2)\,\%$.}
	\label{fig:figure1}
\end{figure}

Figure \ref{fig:figure1}(a) depicts the TMR and TMS minor loops for an MTJ with a nominal barrier thickness of 1.8\,nm MAO. Both loops show identical switching behavior, allowing the identification of clear antiparallel and parallel states. With Eq. $\left(\ref{eq:TMS}\right)$ and a laser power of 150\,mW the TMS amounts to 3.3\,\% while the TMR ratio is 34\,\%. Altogether, the TMS (TMR) ratios are relatively constant with a variation of $\pm$0.25\,\% ($\pm$1\,\%) between different junctions. For this, we measured the TMS (TMR) effect at more than five (ten) junctions. A similar TMS ratio was found for Co$_{40}$Fe$_{40}$B$_{20}$/MgO MTJs \cite{Boehnke1}. The TMR ratio is comparable to similar studies using sputter deposition from a composite, stoichiometric MAO target \cite{tao2}. 

In comparison, Fig. \ref{fig:figure1}(b) displays the TMS and TMR results of the Co$_{25}$Fe$_{55}$B$_{20}$/MgO MTJ. It exhibits an almost rectangular switching behavior resulting in a TMS ratio of (23$\pm$3)\,\% and a high TMR ratio of around 200\,\% indicating very good stack quality. Since the TMS depends on the electronic band structure of the electrodes, both TMS ratios are in good agreement with theoretical predictions for different Co and Fe compositions \cite{stern,heiliger} and experimental results of Co$_x$Fe$_y$/MgO/Co$_x$Fe$_y$ MTJs in case of laser induced heating \cite{walter}.

Figure \ref{fig:figure1}(c) shows the Seebeck voltages of the MTJ with MAO barrier in the parallel and antiparallel magnetization alignment and the corresponding TMS ratios for different laser powers. A barrier of 1.8\,nm MAO results in an averaged (over all laser powers) TMS ratio of $\left(3.3\pm0.2\right)\,\%$. In accordance with previous experiments \cite{walter}, a linear increase of the Seebeck voltages with increasing laser power is observed. 

\subsection{COMSOL simulations}\label{COMSOL}
\begin{figure}[bt]\centering
		\includegraphics{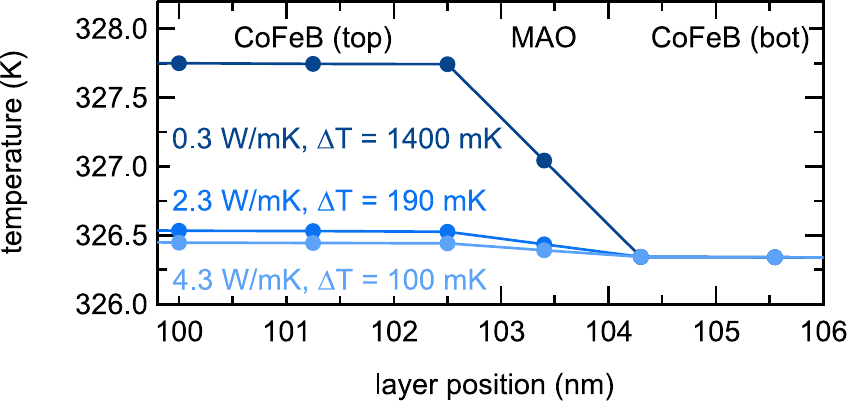}
	\caption{COMSOL simulation of the temperature profile across the tunnel barrier with an applied laser power of 150\,mW (120\,mW at the sample as deduced from calibration measurements) for different thermal conductivities of MAO. The layer position of 0\,nm corresponds to the top of the stack.}
	\label{fig:figure2}
\end{figure}

Simulations are performed with COMSOL Multiphysics to estimate the temperature differences across the barriers and to calculate the Seebeck coefficients. A crucial point within these simulations is the thermal conductivity of thin films as reported in Refs. \onlinecite{mgo} and \onlinecite{drop}. 
For the MTJ with MgO barrier, a value of $\Delta T=$11\,mK was found for a laser power of 150\,mW\cite{Boehnke2} resulting in Seebeck coefficients of $S_{\text{p}}=(-1010\pm20)\,\frac{\mu\text{V}}{{K}}$ and $S_{\text{ap}}=(-1320\pm20)\,\frac{\mu\text{V}}{{K}}$. The values used for the simulation of MTJ with MAO barrier are given in Tab. \ref{comsol} and the results are shown in Fig. \ref{fig:figure2}. Since the thermal conductivity decreases for thin films in comparison to its bulk value and a similar behavior for thin MgO films is observed in Ref. \onlinecite{mgo}, we assume a thermal conductivity of $(2.3\pm2)\,\frac{\text{W}}{\text{m K}}$ for MAO, which is one tenth of the bulk value. Please note that according to Ref. \onlinecite{drop} the thermal conductivity is very sensitive to the imbalance of phonon and electron temperature at nano-magnetic interfaces, which is why we use a large error range for the thermal conductivity of MAO. With the aforementioned assumption and an applied laser power of 150\,mW, the temperature difference across the tunneling barrier varies between 100\,mK and 1400\,mK. The Seebeck coefficients of the MTJ are given by $S_{\text{ap,p}}=-\frac{V_{\text{ap,p}}}{\Delta T}$. Thus, we get: S$^{1.8}_{\text{ap}}\approx-160\,\frac{\mu \text{V}}{\text{K}}$ and S$^{1.8}_{\text{p}}\approx-150\,\frac{\mu \text{V}}{\text{K}}$. With respect to the uncertainty of the thermal conductivity of MAO, an error of $\pm 140\,\frac{\mu \text{V}}{\text{K}}$ is calculated.
\begin{table}[bt]\centering
	\caption{COMSOL simulation parameter values of thermal conductivity $\kappa$, density $\rho$ and heat capacity $C_p$. If not stated otherwise, the values of Walter et al.\cite{walter} are taken. Numbers in rounded brackets are bulk values.}
	\label{comsol}
		\begin{tabular}{c c c c}
		\hline\hline
			material & $\kappa \left(\frac{\text{W}}{\text{m K}}\right)$ & $\rho \left(\frac{\text{kg}}{\text{m}^3}\right)$ & $C_p \left(\frac{\text{J}}{\text{K kg}}\right)$ \\
			\hline
			Ta & 57 & 16650 & 140\\
			Ta$_2$O$_5 ^{\text{a,b}}$ & 0.2 & 8270 & 306\\
			Au & 320 & 19320 & 128\\
			Ru & 117 & 12370 & 238\\
			MnIr$^{\text{c}}$ & 6 & 10181 & 316\\
			CoFeB & 87 & 8216 & 440\\
			MAO$^{\text{d,e,f}}$ & 2.3$\pm2$ (22-24) & 3650 & 815\\
			\hline \hline
		\end{tabular}
		\\
		$^\text{a}$Reference \onlinecite{taox1}. $^\text{b}$Reference \onlinecite{taox2}. $^\text{c}$Reference \onlinecite{mnir}.\\
			$^\text{d}$Reference \onlinecite{mao1}. $^\text{e}$Reference \onlinecite{mao2}. $^\text{f}$Reference \onlinecite{mao3}.\\
\end{table}

\subsection{Brinkman model}
Directly after the TMS measurements, I/V characteristics are recorded at the same junctions. The voltage dependent current density between two electrodes separated by a thin insulating layer is described by Brinkman's model \cite{brinkman}, which is based on the generalized Simmons' model \cite{simmons}. Regarding the relatively low TMR ratio of the MAO MTJ, the presence of a symmetry filter effect can be excluded, which is a basic requirement for the validity of both models. This effect and the accompanying coherent tunnel process is responsible for the high TMR in the CoFeB/MgO MTJs, therefore excluding them from being taken into account within the subsequent Brinkman evaluation. Also, band structure effects caused, for example, by ferromagnetic half-metals are not included in both models. Thus, we focus on the MAO MTJ for the Brinkman evaluation. Within his model, Simmons assumes the potential of the barrier to be symmetric. In order to account for asymmetric barriers, Brinkman replaces the symmetric barrier potential by a trapezoidal barrier potential. The current density (in A/cm$^2$) is then given by
\begin{align}\label{brinki}
J(V)=3.16\cdot10^{10}\,\frac{\varphi^{\frac{1}{2}}}{d}\exp\left(-1.025\,\varphi^{\frac{1}{2}}\,d\right)\cdot\\ \left[V-\frac{A_0\,\Delta\varphi}{32\,\varphi^{\frac{3}{2}}}\,e\,V^2+\frac{3\,A^{2}_{0}}{128\,\varphi}\,e^2\,V^3 \right]
\end{align}
with $A_0=\frac{4\,d\sqrt{2\,m_{\text{eff}}}}{3\,\hbar}$. $\varphi$ is the barrier height (in V), d is the thickness of the barrier (in $\mathring{\text{A}})$, $\Delta\varphi$ is the barrier asymmetry (in V), $e$ is the elementary charge, $\hbar$ is the reduced Planck constant and $m_{\text{eff}}$ is the effective electron mass. Brinkman states that in the case of $\Delta\varphi /\varphi < 1$ and $d > 10\,\mathring{\text{A}}$ the error of this solution amounts to $\leq10\,\%$. The characteristic values of the barrier (height, thickness and asymmetry) are obtained with
\begin{align}\label{eq:brink}
& \varphi^2=\frac{e^2\,C}{32\,A}\,\text{ln}^2\left(\frac{h^3}{\sqrt{2}\,\pi\,e^3\,m_{\text{eff}}}\sqrt{A\,C}\right),\notag \\[2ex]
& d=-\frac{\hbar}{\sqrt{8\,\varphi\,m_{\text{eff}}}}\,\text{ln}\left(\frac{h^3}{\sqrt{2}\,\pi\,e^3\,m_{\text{eff}}}\sqrt{A\,C}\right), \\[2ex]
& \Delta\varphi = -\frac{12\,\hbar}{e\,\sqrt{2\,m_{\text{eff}}}}\frac{\varphi^{\frac{3}{2}}}{d}\,\frac{B}{C}\notag,
\end{align}
where $A$, $B$ and $C$ are the parameters of a second order polynomial fit to the differential conductance given by $dJ/dV=AV^2+BV+C$.

\begin{figure}[bt]\centering
		\includegraphics{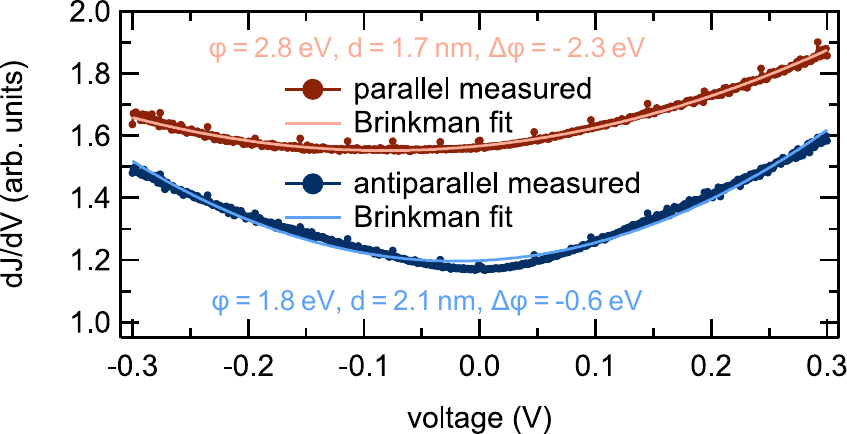}
	\caption{dJ/dV curves (dark) with corresponding Brinkman fits (light) in the antiparallel (blue) and parallel (red) case. In addition, the resulting values for the barrier height $\varphi$, the barrier thickness $d$ and the barrier asymmetry $\Delta\varphi$ are shown.}
	\label{fig:figure3}
\end{figure}

Figure \ref{fig:figure3} shows the results with both dJ/dV curve and Brinkman fit for the antiparallel and parallel magnetization alignment.
Using the Brinkman model, the barrier height $\varphi$, the barrier asymmetry $\Delta\varphi$, and the barrier thickness $d$ are calculated (results are listed in Fig. \ref{fig:figure3}). In the antiparallel case, the change of base temperature of the whole film stack induces a resistance change that is much larger than for the parallel case, which is why the Brinkman fit is not able to cover all features. Still, a good estimation of the barrier parameters is obtained, if compared to the results of the parallel case. It is noteworthy that the theoretically predicted value of the effective electron mass of $m_ {\text{eff}} = 0.422\,m_\text{e}$ \cite{meff} results in a barrier thickness of $d=1.7$\,nm for the parallel case matching the nominal value of 1.8\,nm within the 10\,\% error range of the Brinkman model.

\subsection{Intrinsic TMS \& Brinkman model}

Zhang, Teixeira et al. \cite{Zhang,Teixeira} measured V/I characteristics of CoFeB/MgO MTJs and derived 'intrinsic' Seebeck coefficients via the slope of the symmetric contribution ($(V_++V_-)/2$), where the temperature difference is generated by the Joule heating of the tunnel current. Within this model, the slope of the antisymmetric contribution ($(V_+-V_-)/2$) is directly correlated with the stack resistance. They neglect the general nonlinearity of tunnel processes and the accompanying dependence of the resistance on the voltage.
In order to probe the validity of this intrinsic method, we now compare the results of the laser induced TMS with the intrinsic TMS. 

\begin{figure}[bt]\centering
		\includegraphics{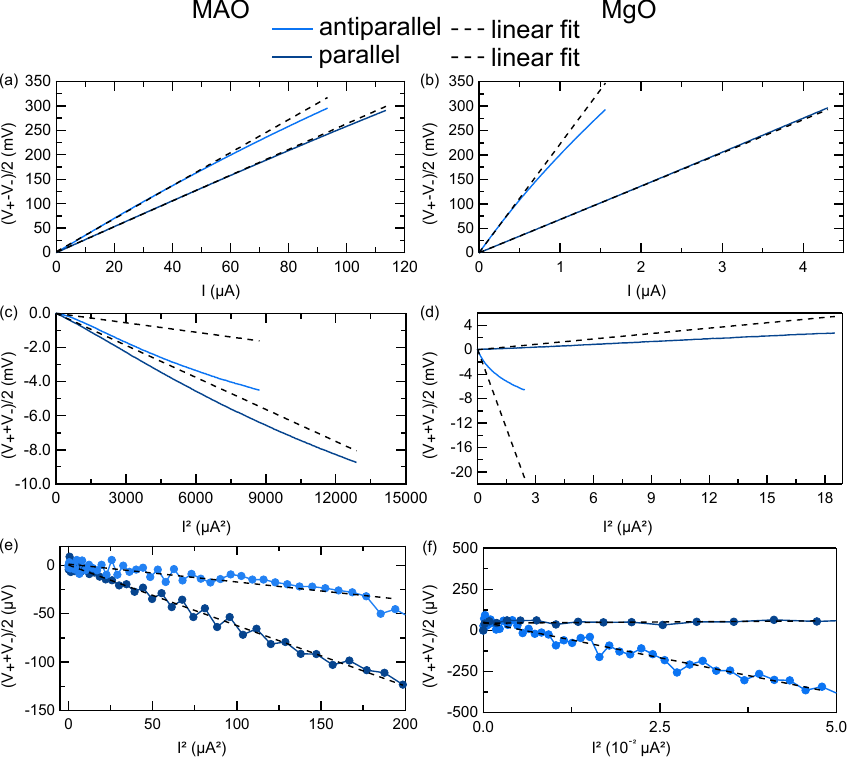}
		\caption{(a),(b) Antisymmetric contribution of the V/I characteristics. The dashed black lines are linear fits and illustrate the non-linearity of the experimental data. (c),(d) Linear fits to the symmetric contributions that are performed within (e),(f) the small and then extended to (c),(d) the whole range.}
		\label{fig:figure4}
		\end{figure}

Figures \ref{fig:figure4}(a),(b) depict the antisymmetric contributions of the V/I-characteristics for both MAO and MgO barrier MTJs. They show a linear increase which diminishes for high currents and both magnetization alignments. This deviation from a purely linear behavior is caused by the changing resistance of the junction due to the induced base temperature changes with increased currents. Thus, additional terms of odd power are present in the I/V data and picked up by the asymmetry evaluation. The deviation from the linear behavior is more prominent in the antiparallel case.
In case of the symmetric parts we find, execpt for the MgO MTJ in the parallel state, negative, non linear contributions [cf. Figs. \ref{fig:figure4}(c),(d)]. Therefore, only the first part of the curves is fitted to extract the Seebeck coefficients via the method of the intrinsic TMS [cf. Figs. \ref{fig:figure4}(e),(f)]. 

In addition and to follow the method proposed by Zhang, Teixeira et al.  \cite{Zhang,Teixeira}, we need to calculate $\sum\limits_j \eta_j\,R_j\,R_{\kappa_j}$, where $\eta_j$ is the thermal asymmetric parameter obtained via $\eta_j=\frac{\nabla T\,\kappa_j\,\sigma_j}{\text{J}^2\,\text{d}^2_j}$ , $R_j$ is the resistance and $R_{\kappa_j}=\frac{d}{\kappa_j\,A}$ is the heat resistance. $\nabla T$ refers to the temperature gradient across the barrier, $\kappa$ to the thermal conductivity, $A$ to the area of the junction, $\sigma$ to the electric conductivity and $J$ to the current density given by $J=I/A$ ($I$:\,current, $A$:\,area). The index $j$ identifies each individual layer.
Since the stacks show resistances of several k$\Omega$ originating mostly from the MAO and MgO barrier, we neglect the influences of other layers and assume $\sum\limits_j \eta_j\,R_j\,R_{\kappa_j}=\eta_{\text{MAO/MgO}}\,R_{\text{MAO/MgO}}\,R_{\kappa_{\text{MAO/MgO}}}=\alpha$ with $R_{\text{MAO/MgO}}=R$ the resistance of the stack. From the TMR measurements we obtain R$_{\text{ap}}$ (R$_{\text{p}})=3.6\,\text{k}\Omega$ $(2.7\,\text{k}\Omega$) for the MAO barrier and R$_{\text{ap}}$ (R$_ {\text{p}})=195\,\text{k}\Omega$ ($66\,\text{k}\Omega$) for the MgO barrier. 

With these values and the thickness of the barrier, the area of the junction and the thermal conductivity of MAO and MgO mentioned in section\ref{COMSOL} , $\sigma_{\text{MAO/MgO}}$ and $R_{\kappa_{\text{MAO/MgO}}}$ are calculated. Additionally, we take a mean current resulting from the I/V curves of $50\,\mu$A for the MAO barrier and $1\,\mu$A for the MgO barrier  and, furthermore, assume a temperature gradient of $25\,\frac{\text{mK}}{\text{nm}}$ for MAO and $1\,\frac{\text{mK}}{\text{nm}}$ for MgO (please note: Teixeira et al. used a temperature gradient of $75-195\,\frac{\text{mK}}{\text{nm}}$ generated by a current of 0.4\,mA). Thus, we get values for the symmetric slope as well as for the $\alpha$ parameter, which are summarized in Tab.\ref{evaluation} and which allow to directly compare the results of the intrinsic TMS with the results of the laser induced TMS [cf. Tab. \ref{intrinsic2}].

\begin{table*}[bt]\centering
	\caption{Results of the intrinsic symmetry evaluation. The resistances $R$ extracted from the TMR loops are given as a comparison to the resistances from the antisymmetric contributions. A satisfying agreement between the two methods is achieved within the measurement uncertainty. Additionally, the large difference of the resistances between the MTJ with MAO and MgO barrier is also found in the slope of the symmetric contributions. After rounding to two significant digits, no difference remains in the $\alpha$ parameter for the parallel and antiparallel states of both MTJs.}
	\label{evaluation}
	\begin{tabular}{c c c c c}
	\hline\hline
			sample [state] & antisymmetric slope (k$\Omega$) & R from TMR (k$\Omega$) & symmetric slope (V/A$^2$) & $\alpha$ (K/A$^2$)  \\
			\hline
			MAO [p] & $2.6\pm0.1$ & $2.7\pm0.1$ & $-6.3\cdot10^5\pm10^4$ & $5.4\cdot10^{14}$ \\
		  MAO [ap] & $3.4\pm0.1$ & $3.6\pm0.1$ & $-1.8\cdot10^5\pm10^4$ & $5.4\cdot10^{14}$ \\
			MgO [p] & $68\pm2$ & $66\pm2$ & $2.9\cdot10^8\pm2.5\cdot10^8$ & $1.2\cdot10^{15}$ \\
		  MgO [ap] & $223\pm10$ & $195\pm20$ & $-8.6\cdot10^9\pm3\cdot10^8$ & $1.2\cdot10^{15}$  \\
			\hline \hline
	\end{tabular}
\end{table*}

\begin{table}[bt]\centering
	\caption{Results of the intrinsic TMS evaluation. The results of the laser induced TMS are given as a direct comparison. The error ranges of the intrinsic Seebeck coefficients result from the inaccuracy of the linear fits to the symmetric contributions.}
	\label{intrinsic2}
	\begin{tabular}{c c c c }
	\hline\hline
			& $S_{\text{p}}\,\left(\frac{\text{$\mu$V}}{\text{K}}\right)$  &  $S_{\text{ap}}\,\left(\frac{\text{$\mu$V}}{\text{K}}\right)$ & TMS\,$\left(\%\right)$
			\\
			\hline
			MAO MTJ& & & \\
			\\
			intrinsic&$-1.2\cdot10^{-3}\pm10^{-4}$&$-3\cdot10^{-4}\pm10^{-4}$& $-75\pm10$ \\
			\\
		  laser&$-150\pm140$&$-160\pm140$&$3.3\pm0.2$ \\
			MgO MTJ& & & \\
			\\
			intrinsic&$0.3\pm0.2$&$-7.5\pm0.2$&$104\pm3$ \\
			\\
		  laser&$ -1010\pm20$& $-1320\pm20$&$23\pm3$ \\
			\hline \hline
	\end{tabular}
\end{table}

Clearly, the obtained values for the intrinsic Seebeck coefficients do not match with the results of the laser induced TMS, neither for the MAO nor for the MgO barrier. Furthermore, the intrinsic TMS ratios do not coincide with the results of the laser induced TMS ratios. Please note, that changing the aforementioned assumptions only results in different values for the intrinsic Seebeck coefficients. However, the sign of the intrinsic TMS ratio is dominated by the slope of the symmetric contribution. Accordingly, in our case, S$_{\text{p}}$ will always be larger than S$_{\text{ap}}$ for the MTJ with MAO barrier, thus, resulting in a negative TMS ratio. In addition, the Seebeck coefficients obtained from the intrinsic method of the MTJ with MgO barrier show a different sign that is not observed with the laser induced TMS. These findings directly contradict the results of the laser induced TMS. Therefore, we are not able to identify any reasonable contribution of the intrinsic TMS which would be comparable to the more clear laser induced TMS.
\begin{figure}[bt]\centering
		\includegraphics{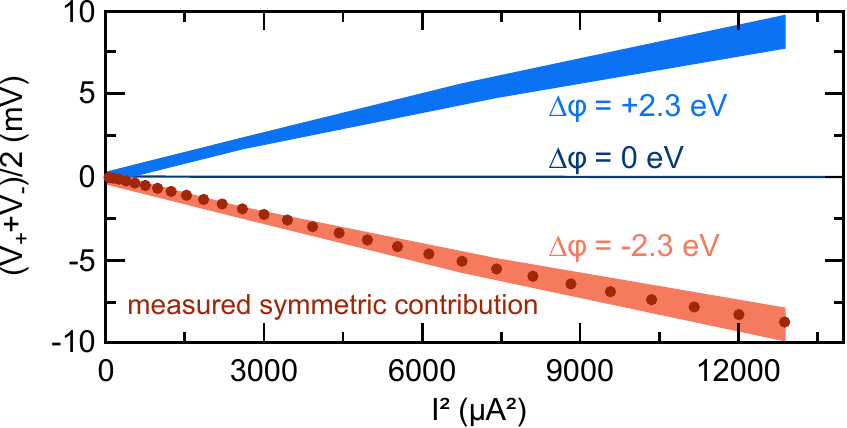}
	\caption{Original symmetric contribution of the MAO MTJ in the parallel case (dark red), corresponding Brinkman fit ($\Delta\varphi=-2.3\,$eV) (light red), simulated barrier asymmetry of $\Delta\varphi=+2.3\,$eV (light blue) and $\Delta\varphi=0\,$eV (dark blue). For the sake of clarity, only one in ten data points of the original data is shown. The colored areas represent the typical error range of the Brinkman model of 10\,\%.}
	\label{fig:simulation}
\end{figure}

However, the Brinkman model offers an alternative way to explain the occurring antisymmetric and symmetric contributions in case of the MAO MTJ. We focus on the parallel case where a good agreement between data and model is obtained [cf. Fig. \ref{fig:figure3}]. Now, the symmetry evaluation is performed with simulated I/V curves based on the Brinkman model with different values for the barrier asymmetry $\Delta\varphi$. Figure \ref{fig:simulation} shows the results of the symmetry evaluation of the original data, its corresponding Brinkman fit, a reversed barrier asymmetry and a vanishing barrier asymmetry. Obviously, the barrier asymmetry plays a vital role for the symmetric contribution of the V/I curve.

Please note that a symmetric barrier shows no symmetric contribution, making the identification of an intrinsic TMS impossible. In contrast, the asymmetric contributions are the same for different values of the barrier asymmetry. Thus, the symmetric contribution of the V/I curve in the parallel case is very well described by the Brinkman model even without any assumptions of temperature differences.
\section{Conclusion}
We have investigated the TMS effect of Co$_{40}$Fe$_{40}$B$_{20}$/MAO and Co$_{25}$Fe$_{55}$B$_{20}$/MgO MTJs with laser induced heating. In case of a barrier consisting of MAO, the TMS ratio of about $3\,\%$ as well as the Seebeck coefficients are consistent with findings of other groups who used similar materials. The results of the MgO based MTJs show large TMR ratios of up to $200\,\%$ and TMS ratios of around $20\,\%$. This TMS ratio is directly related to the different CoFeB composition. 
In addition, we have studied the symmetry of I/V characteristics within the framework of the intrinsic TMS proposed by Zhang, Teixeira et al. Both, antisymmetric and symmetric contributions, revealed deviations from the expected linear behavior suggested by the model of the intrinsic TMS. Our findings show that it is not possible to consistently compare the results of laser induced and intrinsic TMS. Nevertheless, the Brinkman model offers an alternative way to explain the occurring features if no $\Delta_1$ symmetry filter effect is present, giving a consistent explanation for the observed symmetric contribution. In particular, we found that the symmetric I/V contribution in the parallel case is solely influenced by the barrier asymmetry. Thus, we conclude that the symmetry analysis is not suitable to unambiguously identify an intrinsic TMS.
\section{Acknowledgments}
The authors gratefully acknowledge financial support by the Deutsche Forschungsgemeinschaft (DFG) within the priority program Spin Caloric Transport (SPP 1538). They further thank Christian Heiliger from the University of Gie\ss en for theoretical input.

\end{document}